# Experimental demonstration of time coding quantum key distribution

**William Boucher and Thierry Debuisschert**
*THALES TRT, Domaine de Corbeville, 91404 Orsay Cedex, France*
thierry.debuisschert@thalesgroup.com

**Abstract :** Time coding quantum key distribution with coherent faint pulses is experimentally demonstrated. A measured 3.3 % quantum bit error rate and a relative contrast loss of 8.4 % allow a 0.49 bit/pulse advantage to Bob.
© 2005 Optical Society of America

Quantum key distribution has been widely developed in recent years [1]. In view of practical applications we have proposed a simple protocol based on time coding [2]. A related protocol has been proposed recently using phase coherent pulses produced with a mode-lock laser [3]. We describe here an experimental realization of our protocol.

Time coding makes use of coherent one photon pulses with square profile and duration T. The security relies on the minimum time–frequency uncertainty product. In the simplest protocol, two kinds of pulses (b and c) are sent by Alice with a delay of 0 (bit 0) or T/2 (bit 1) with respect to a clock [2], (Figure 1). The detection by Bob may occur in three successive time slots (3, 4 and 5) of duration T/2. An eavesdropper may induce errors if he resends a pulse of duration T after detection in time slot 4 that leads to an ambiguous result. In order to detect a possible shortening of the pulses resent by Eve, half of the pulses received by Bob are sent at random to a Mach-Zender interferometer having a path difference of cT/2. This allows a pulse duration measurement thanks to a contrast measurement [2]. The other half of the pulses received by Bob are sent to a photon-counter to establish the key. A four states protocol with two additional pulses (a and d) carrying no information imposes additional symmetry constraints to Eve, which keeps her to exploit the losses of the quantum channel [2]. Simultaneous measurement of the quantum bit error rate (QBER) on the sifted key and of the interferometer contrast allows to evaluate the security of the transmission. The mutual informations between Alice and Eve ($I_{AE}$) and between Alice and Bob ($I_{AB}$) can be calculated as a function of the QBER, the contrast being a parameter [2]. The criterion $I_{AB} \geq I_{AE}$ allows the determine the parameters values for which the security can be guaranteed.

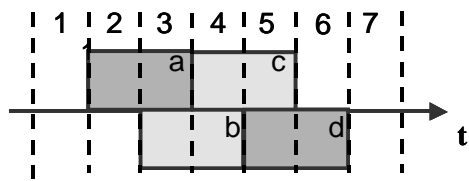

Figure 1 : Principle of the time coding protocols. In the two states protocol, only pulses b and c, corresponding to bit 0 and 1 respectively, are emitted by Alice. In the 4 states protocols, two additional pulse a and d carrying no information can be launched by Alice.

In the experimental set-up (Figure 2), an electrical pulse generator drives an electro-optic modulator that tailors the pulses in the beam produced by a cw diode laser emitting at 852 nm. The wavelength is chosen mainly for the availability of efficient Si photon-counters in the infrared. The temporal resolution of the photon-counters is 300 ps. Thus ideal pulses would be chosen with a duration of 20 ns to get sufficient resolution. Due to the imperfection of the electrical generator (3 ns rise-time and decay-time), the actual width is 18.7 ns (FWHM). The pulse are then strongly attenuated down to an average 0.1 photon per pulse and sent to Bob trough an optical fiber. At the entrance of Bob set-up, a 50/50 beamsplitter sends at random the pulses to the photon-counter used to establish the key or to the Mach-Zender interferometer (path difference of 3 m). Two photon-counters with balanced efficiencies measure the number of photons in each output port.

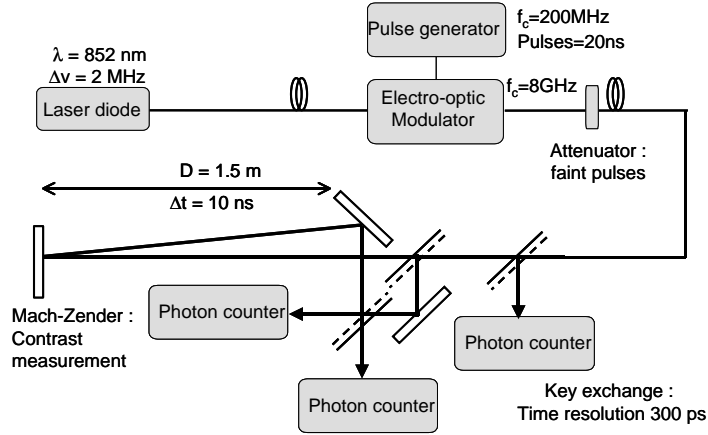

Figure 2 : Experimental set-up. The pulses received from Alice are directed by bob either to a Mach-Zender interferometer to measure their average duration or to a photon counter to establish the key.

Alice sends a sequence of pulses with a possible delay of 10 ns with respect to a reference clock which is common to Bob. The clock period is 100 ns. The duration of the sequences sent by Alice is 3.2 ms. The sequence duration is determined by the memory depth of the digitizing oscilloscope used to register the chronogram of the photodetection events. Due to the actual profile of the pulses, the expected contrast is 57.6 % for a perfectly coherent source instead of 50 % for perfect 20 ns squares. The actual value measured with intense pulses is 54 % resulting in a relative contrast loss of 6.1 %.

To evaluate the QBER, Alice sends 3.2 ms long sequences of pulses with alternate delays 0 ns (bit 0) and 10 ns (bit 1). The attenuator is tuned in order to get 0.1 photon per pulse in average. The error is the ratio between the number of detections in the wrong time-slot (e.g time-slot 5 for bit 0) to the total number of non ambiguous detections (time-slots 3 and 5). The QBER value is found by post processing the data finding the minimal error value as a function of the propagation delay between Alice and Bob.

The second parameter that has to be evaluated is the average autocorrelation function of the pulses received by Bob for a delay of 10 ns denoted $\gamma$. In the photon counting regime, the contrast of the interferometer is measured by the difference between the output photon numbers normalized to their sum. The contrast, denoted C, depends of two random variables. The first one is the phase $\phi$ of the interferometer which is let as a free-running parameter with no control-loop between Alice and Bob for practical reasons. The second one is the photon number in the output ports of the interferometer which is affected by shot-noise. For a given value of the autocorrelation function $\gamma$ and of the phase, the contrast can take values characterized by a density probability which is a gaussian centered on $\gamma \cos(\phi)$ with a variance $\sigma^2 = 1/N_p$ where $N_p$ is the average number of photons in each sequence. In order to improve the accuracy in the evaluation of $\gamma$, measurements over $N_s$ successive sequences are performed. From the $N_s$ measured values of the contrast, one can deduce the probability density of $\gamma$. It is a gaussian centred on $\gamma_0$ given by $\gamma_0^2 = 2\left(\overline{C^2} - \sigma^2\right)$ where $\overline{C^2}$ is the variance of the contrast measurements (assumed centred). The variance of the corresponding distribution is given by $\sigma_T^2 = 2/N_p N_s$. It shows that the total number of detected photons is taken into account resulting in an increase in the precision measurement of $\gamma$. The factor 2 is a correction resulting from the average on squared cosine of the random phases. Knowing the distribution of $\gamma$, Bob can evaluate the minimum value of the pulses autocorrelation for a given level of security. For example, if he tolerates an error probability of 1e-3, the minimum value of $\gamma$ is not smaller than $\gamma_0 - 3\sigma_T$. Then, from the corresponding $I_{AE}$ curve on the security diagram [2] (figure 3), Bob can either calculate is information advantage on Eve for the measured QBER or evaluate the maximum QBER allowing security from the relation $I_{AB}=I_{AE}$.

Applying that method, we have registered 290 successive sequences with an average of 282.5 photons per sequence. The variance of the measured contrast distribution is equal to $\overline{C^2} = 0.15$. The sequence noise variance is equal to $\sigma^2 = 3.54e-3$. The central value of the autocorrelation distribution is thus equal to $\gamma_0 = 0.541$. The value is the same as that measured with intense pulses. With our experimental data, $\sigma_T$ is equal to 4.9e-3, which leads to $\gamma_{3\sigma} = 0.526$. From that value, one can calculate the contrast loss relative to the theoretical value defined by $dC = (\gamma_{th} - \gamma)/\gamma_{th}$. The theoretical value calculated from the actual shape of our pulses is $\gamma_{th} = 0.576$. With $\gamma$ equal to $\gamma_{3\sigma}$, the resulting value of $dC$ is 8.6 %. Taking only into account the imperfections of the set-up or for an infinite number of photons, $\gamma$ is equal to $\gamma_0$ and the resulting value of $dC$ is 6.1 %. It shows that the statistical noise it not dominant. It could in addition be reduced up to being negligible increasing the number of sequences.

In the same series of measurements, we have measured the QBER of the last sequence. We have found a value of 3.3 %. It shows clearly that small QBER can be obtained even if the pulses do not have a perfect profile. With such a value, the information advantage of Bob on Eve is 0.49 bits/pulse. With no statistical noise, it would be 0.54 bits/pulse. In the case of a perfect system (dC=0) it would be 0.72 bits/pulse. For those three curves one can calculate the maximum QBER value given by the relation $I_{AB}=I_{AE}$. One finds respectively 9.7 %, 11 % and 17 %, which shows that the set-up can accommodate higher QBER than that experimentally measured.

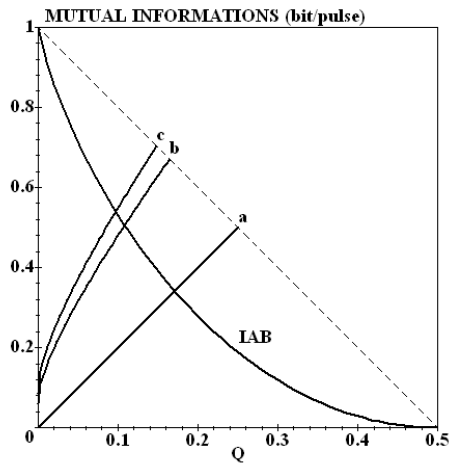

Figure 3 : Mutual informations as a function of the QBER (Q). $I_{AB}$ decreases with Q and is independent of the contrast. IAE increases with Q and is given for three values of the relative contrast loss : 0 % ; 6.1 % and 8.4 % (a, b and c respectively).

That experiment is a demonstration of principle that validates the time coding for quantum key distribution. A natural extension would be to develop a prototype working at 1550 nm for telecom applications. The typical loss of 0.2 dB/km would result in an increase in the secure transmission range. Taking into account the characteristics on InGaAs photon-counters at 1550 nm (efficiency 10 %; dark-counts $10^5$ / s) leads to typical secure range of 20 km in the case of intercept-resend attacks [2]. Such a range combined with the great simplicity of the prototype makes time coding a valuable candidate for a practical implementation of quantum key distribution.